# Cross-spectrum Measurement of Thermal-noise Limited Oscillators

A. Hati, C. W. Nelson and D. A. Howe
Time and Frequency Division
National Institute of Standards and Technology
Boulder, CO/USA

*Abstract:* Cross-spectrum analysis is a commonly-used technique for the detection of phase and amplitude noise of a signal in the presence of interfering noise. It extracts the desired correlated noise from two time series in the presence of uncorrelated interfering noise. Recently, we demonstrated that the phase-inversion (anti-correlation) effect due to AM noise leakage can cause complete or partial collapse of the cross-spectral function [1], [2]. In this paper, we discuss the newly discovered effect of anti-correlated thermal noise that originates from the common-mode power divider (splitter), an essential component in a cross-spectrum noise measurement system. We studied this effect for different power splitters and discuss its influence on the measurement of thermal-noise limited oscillators. An oscillator whose thermal noise is primarily set by the 50 Ω source resistance is referred to as a thermally-limited oscillator. We provide theory, simulation and experimental results. In addition, we expand this study to reveal how the presence of ferrite-isolators and amplifiers at the output ports of the power splitters can affect the oscillator noise measurements. Finally, we discuss a possible solution to overcome this problem.

## I. INTRODUCTION

Oscillators enable much of our modern technology, including smart phones, GPS receivers, radar/surveillance/imaging systems, electronic test and measurement equipment, and much more. System designers and manufacturers need oscillators with the lowest possible phase noise (timing jitter or spectral purity), especially for high performance applications. However, the measurement of phase noise of many state-of-the-art oscillators at current noise levels is challenging; commercial phase noise measurement systems give results varying by more than a factor of 10, often severely under-reporting phase noise [3].

The cross-spectrum technique is a common tool used for the measurement of low-phase and amplitude noise oscillators [4]–[14]. It uses two independent channels as discussed in Section II; each consists of a reference oscillator and a phase detector (PD) that simultaneously measures the noise of the oscillator under test. Computing the cross-spectral density of voltage fluctuations between two channels improves the spectral resolution of the noise measurements by reducing the effect of uncorrelated noise sources in each channel by $\sqrt{m}$, where $m$ is the number of averages of the fast Fourier Transform (FFT). If two channels are statistically independent, the average cross spectrum converges to the DUT noise spectrum. Until very recently, it was believed that the cross-spectrum method always over-estimates the measurement of DUT noise in the presence of correlated but unwanted and uncontrolled noise phenomena affecting both channels (DUT AM noise, vibration induced noise, EMI etc.) However, it was demonstrated in [1] and [2] that if two time series, each composed of the summation of two fully independent signals, are correlated in the first time signal and anti-correlated (phase inverted) in the second, and have the same average spectral magnitude, the cross-spectrum power density between two time series is annihilated and collapses to zero. This effect can lead to dramatic under-reporting of the DUT noise. These conditions may occur only at localized offset frequencies or over a wide range of frequency of the cross-spectrum. Significant partial annihilation can occur if the interfering noise is within 10 dB of the desired noise. Such interfering signals can either be correlated to the DUT or completely uncorrelated. In [2], the anti-correlation collapse mainly due to AM noise leakage was discussed. More recently a different source of anti-correlation in a cross-spectrum measurement has been identified; the origin is from the common-mode power splitter (reactive Wilkinson or resistive). The correlated thermal noise of the power splitter appears equally but in opposite phase in two channels of the cross-spectrum system. This new source of phase-inverted interfering noise was first addressed by Joe Gorin [15]. As early as the year 2000, anomalously low-noise in a cross-spectrum measurement system was reported by Ivanov and Walls and it was interpreted that it is possible to measure the additive noise of a device with an effective temperature much lower than the ambient temperature [16], [17]. Those results were in reality an observation of anti-correlated cross-spectrum thermal noise measurements. In this paper, we will discuss the influence of anti-correlated thermal noise of the power splitter on the thermally-limited oscillator noise measurements. When we say thermally-limited, we mean that the white signal to noise ratio of the oscillator is at or near the level generated by the thermal noise of a 50 Ω source resistor. We will provide theory, simulation and experimental results and also discuss solutions to overcome this problem.

## II. BRIEF OVERVIEW OF COLLAPSE OF THE CROSS-SPECTRUM

A detailed theory and simulations of the positive correlation and anti-correlation (collapse) of the cross-spectral function are discussed in [12] and [2]. In this section, we briefly revisit these two cases of cross-spectrum. Let us first consider two signals $x(t)$ and $y(t)$, each composed of four statistically independent, ergodic and random processes $a(t)$, $b(t)$, $c(t)$ and $d(t)$ such that

$$\begin{aligned} x(t) &= a(t) + c(t) + d(t), \\ y(t) &= b(t) + c(t) + d(t). \end{aligned} \quad (1)$$

Here, $c(t)$ and $d(t)$ are the desired signals that we wish to recover, and $a(t)$ and $b(t)$ are the uncorrelated interfering signals. If $d(t)$ is correlated in both $x(t)$ and $y(t)$, then the cross-spectrum, $S_{yx}(f)$ converges to the sum of the average power spectral densities (PSD) of $c(t)$ and $d(t)$

$$\langle S_{yx}(f) \rangle = \tfrac{1}{T}\left[\langle CC^*(f) \rangle + \langle DD^*(f) \rangle\right] = S_c(f) + S_d(f). \quad (2)$$

The cross-spectrum $S_{yx}(f)$ is calculated from the ensemble average of the product of truncated Fourier transform of time series $x(t)$ and complex conjugate of Fourier transform $y(t)$. $T$ is the measurement time normalizing the PSD to 1 Hz. However, when

$c(t)$ is correlated in $x(t)$ and $y(t)$ and $d(t)$ is anti-correlated (phase inverted) in $x$ and $y$ as in (3) such that

$$x(t) = a(t) + c(t) + d(t),$$
$$y(t) = b(t) + c(t) - d(t), \quad (3)$$

then the corresponding cross-PSD is represented as

$$\langle S_{yx}(f) \rangle = \tfrac{1}{T}\left[\langle CC^*(f) \rangle - \langle DD^*(f) \rangle\right] = S_c(f) - S_d(f). \quad (4)$$

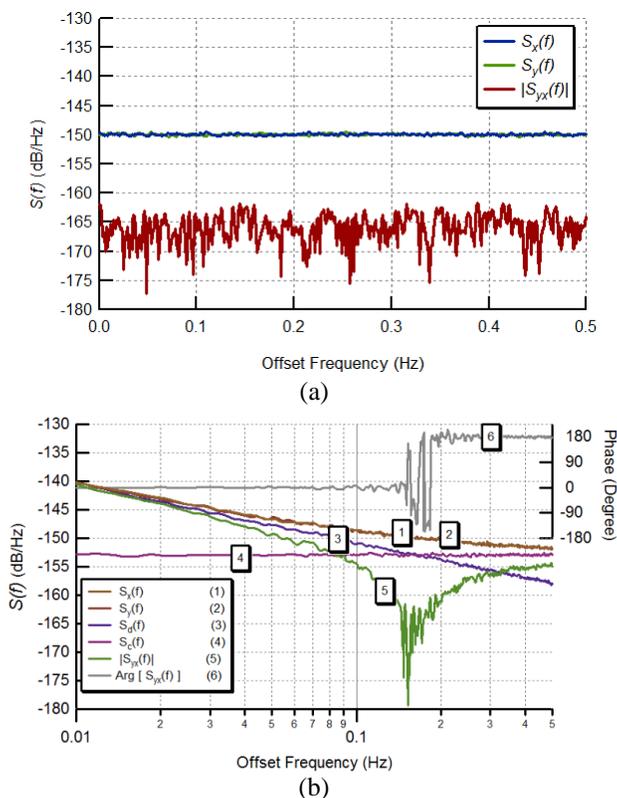

Fig. 1: Mathworks simulation results of the cross-spectrum collapse when $x(t) = c(t) + d(t)$, and $y(t) = c(t) - d(t)$: (a) for the addition of two completely independent noise sources, $c(t)$ and $d(t)$, each with power spectral density of -153 dB/Hz relative to unity. (b) for two independent noise sources, $c(t)$ and $d(t)$, with different frequency dependence are added. Signal $S_c(f)$ has a power spectral density of −153 dB/Hz relative to unity. Signal $S_d(f)$ has a $f^{-1}$ slope and intersects signal $S_c(f)$ at a frequency of 0.16 Hz.

The cross-spectrum in (4) collapses to zero when $c(t)$ equals $d(t)$. In this paper we will mainly discuss the noise measurement conditions that are of type (4). Mathworks Simulink simulation results for two different categories of the cross-spectrum collapse are depicted. Beginning with Fig. 1a, a collapse over a wide range of offset frequencies occurs when two completely independent white noise sources, $c(t)$ and $d(t)$, each with equal power spectral density, are anti-correlated in $x(t)$ and $y(t)$. Second, a localized collapse occurs (Fig. 1b) due to the interaction of two different sloped noise types, this appears as a notch in the magnitude of the cross-spectrum as well as $180^0$ change in its argument. For this simulation we use the biased magnitude estimator $\left|\langle S_{yx}(f) \rangle_m\right|$, as well as its pair argument $\{\langle S_{yx}(f)\rangle_m\}$ for describing the amplitude and phase relationships [12].

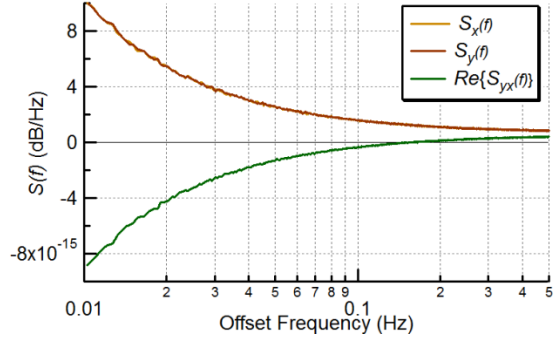

Fig. 2: Plot of $\Re\{\langle S_{yx}(f)\rangle_m\}$ when two independent noise sources, $c(t)$ and $d(t)$, with different frequency dependence are added.

The detection of a cross-spectrum collapse is difficult when noise slopes of the desired and interfering signals are the same. However, in a case when two noise types intersect; the appearance of a notch in the magnitude of the cross-spectrum is a clear indication of the problem. This notch will also have an associated phase change of 180 degrees in the argument of the cross-spectrum as in Fig. 1b. In rectangular coordinates, this will be observed as a change in sign of the real part of the cross-spectrum, $\Re\{\langle S_{yx}(f)\rangle_m\}$ as shown in Fig. 2.

### III. CHALLENGES OF CROSS-SPECTRUM NOISE MEASUREMENT OF A THERMALLY-LIMITED OSCILLATOR

The configuration of a cross-spectrum phase noise measurement is shown in Fig. 3. Here the component noise contributions from each parallel signal path within the dotted dashed blue box appear uncorrelated and are rejected by the cross-spectrum while the noise contributions of components in the red box appear correlated and are retained in the output of the cross-spectrum. In addition to the DUT noise, thermal noise of the power splitter (PS in Fig. 3) will also be correlated in both channels. We will show later that the noise from the green dotted boxes with question mark and the PDs can appear as correlated if the PS doesn't have high isolation between the two outputs. The measurement of white PM or AM noise of most typical oscillators is not near the thermal limit and therefore not significantly biased by the thermal noise of the common-mode-power splitter. However, recently several commercial ultra-low phase noise (ULPN) oscillators have been introduced that are now reaching the thermal limit. In this new class of oscillators, the bias, either positive or negative, from the power splitter thermal noise plays a dominant role. Repeatable and reproducible noise measurements of these ultra-low noise thermally-limited oscillators have become difficult due to the effect of anti-correlated thermal noise originating from the power splitter. An example of this problem is demonstrated in Fig. 4; where the shaded band between the red and black curves represents the range of different phase noise measurement results of the same ULPN oscillator at 100 MHz. Measurement are made using the cross-spectrum technique; each with a slightly different configuration or components. For instance, the phase noise is measured either with different power splitters types (such as Wilkinson, resistive 2-R or 3-R) or different components between the power splitter and the phase detector (such as attenuator/isolator/amplifier). In addition to the thermal noise of the power splitter, the results shown in Fig. 4 might also have been affected by uncontrolled AM noise leakage, ground loops or electromagnetic interference (EMI). Good metrology relies on method validation; the results of differently calibrated methods and measurement configurations should match within the measurement uncertainty [18], [19]. Fig. 4 clearly shows the results varying by more than a factor of 10, either over or severely under-reporting the measured phase noise.

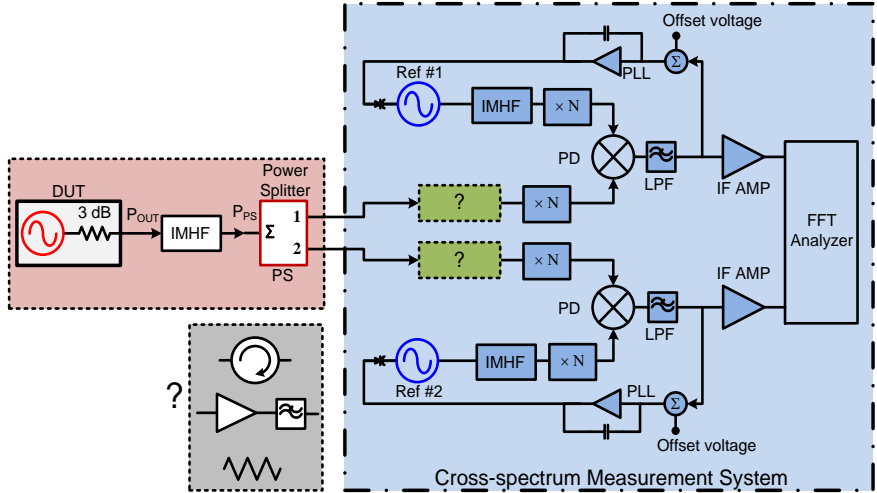

Fig. 3: Block diagram of a cross-spectrum phase noise measurement system. The green dotted box with "**?**" in each channel contains any one component of three shown inside the gray dotted box. The frequency multiplication factor 'N' $\geq 1$. IMHF – Impedance matching and harmonic filtering, LPF – Low pass filter, PD – Phase Detector, FFT- Fast Fourier Transform, PLL – Phase Locked Loop

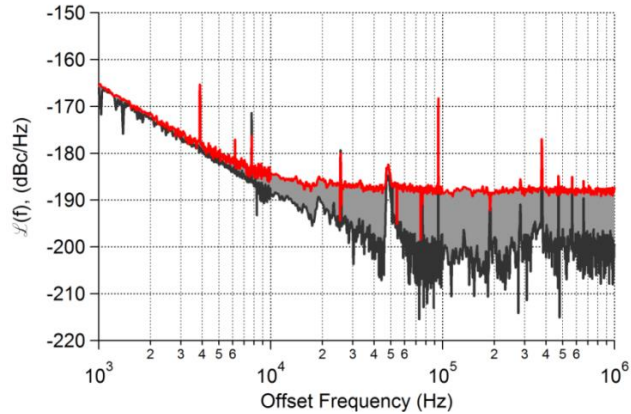

Fig. 4: Variation in the phase noise of a 100 MHz thermally-limited oscillator measured with cross-spectrum system. More than 10 dB difference in the phase noise is observed either by changing the common-mode power splitter type or the measurement configuration. The bottom noise plot is limited by the number of FFT averages $m$; for offset frequencies above 10 kHz, $m$ = 100,000. The bottom black curve and the top red curve correspond to a Wilkinson power splitter with and without the isolation resistor ($R_i$) respectively.

## IV. EFFECT OF POWER SPLITTER THERMAL NOISE ON THE CROSS-SPECTRUM MEASUREMENT

In the following sub-sections, we primarily discuss the theory and simulation studies on the effect of thermal noise of various power-splitter types on the cross-spectrum analysis. Theoretical findings are supported with experimental results.

### A. Theory

The schematic representation of a few power splitters [20]–[22] such as the conventional Wilkinson power splitter (CWPS), the modified Wilkinson power splitter (MWPS), and the resistive 3-R, 2-R and 1-R power splitter configurations are shown in Fig. 5a, b, c, d, e and f, respectively. For the ideal case, the insertion loss in both types of Wilkinson power splitters (Fig. 5a and b) is 3 dB and isolation of each is infinite. On the other hand, for the resistive power splitters 3-R, 2-R and 1-R the loss is 6 dB and the corresponding isolation is 6 dB, 12 dB, and 2.5 dB, respectively. Also, a terminated 3-R splitter (Wye or Delta) presents a 50 Ω impedance looking into any of the three ports. The 2-R and 1-R power splitters both have 50 Ω input impedances, while presenting 83.33 Ω and 30 Ω impedances, respectively, at their output ports.

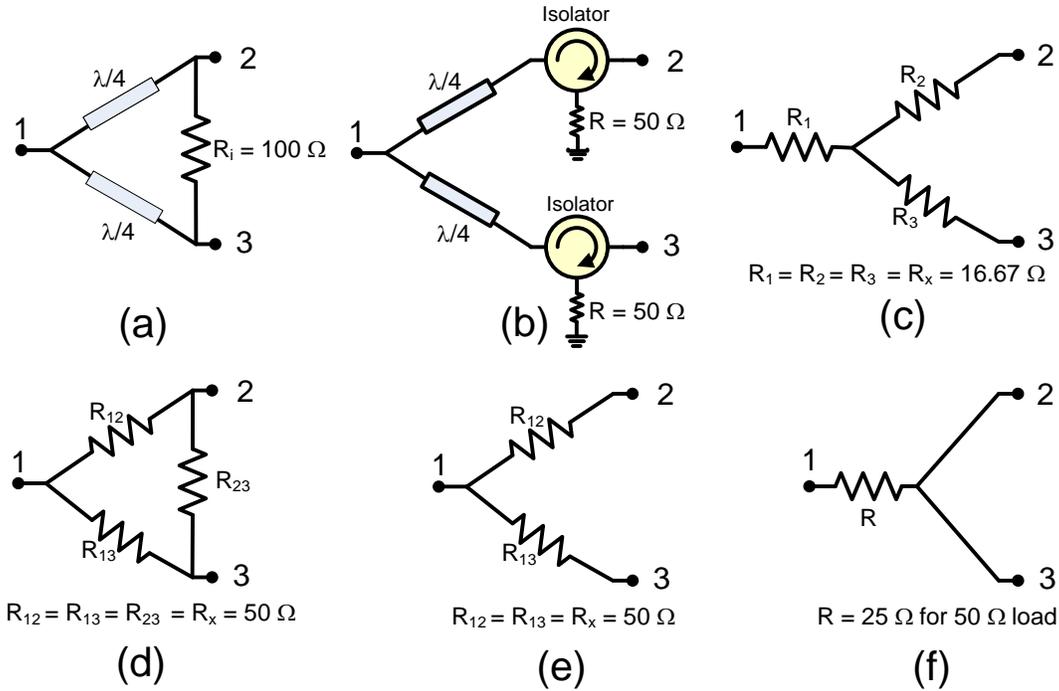

Fig. 5: (a) Conventional Wilkinson power splitter, (b) Modified Wilkinson power splitter, (c) Resistive 3-R (Wye configuration), (d) Resistive 3-R (Delta configuration), (e) Resistive 2-R and (f) Resistive 1-R.

For the analysis of thermal noise of power splitters, we will consider the 3-R power splitter in delta configuration since it closely resembles the CWPS. The equivalent circuit to the delta 3-R power splitter with thermal voltage noise sources for each resistor is shown in Fig. 6. The voltage source $V_S$ corresponds to the DUT source noise, and

$V_{n12}$, $V_{n13}$, $V_{n23}$, $V_{nL1}$ and $V_{nL2}$ are respectively the thermal noise of the power splitter and the load resistances $R_{12}$, $R_{13}$, $R_{23}$, $R_{L1}$ and $R_{L2}$. Assuming a 50 Ω system, all resistors in Fig. 6 are equal to 50 Ω and the corresponding node voltages $v_1$, $v_2$, and $v_3$ at port 1, 2 and 3 can be written as

$$v_1 = \frac{2v_S + v_{n12} + v_{n13} + v_{nL2} + v_{nL3}}{4},$$

$$v_2 = \frac{v_S - v_{n12} + v_{n23} + 2v_{nL2} + v_{nL3}}{4}, \quad (5)$$

$$v_3 = \frac{v_S - v_{n13} - v_{n23} + v_{nL2} + 2v_{nL3}}{4}.$$

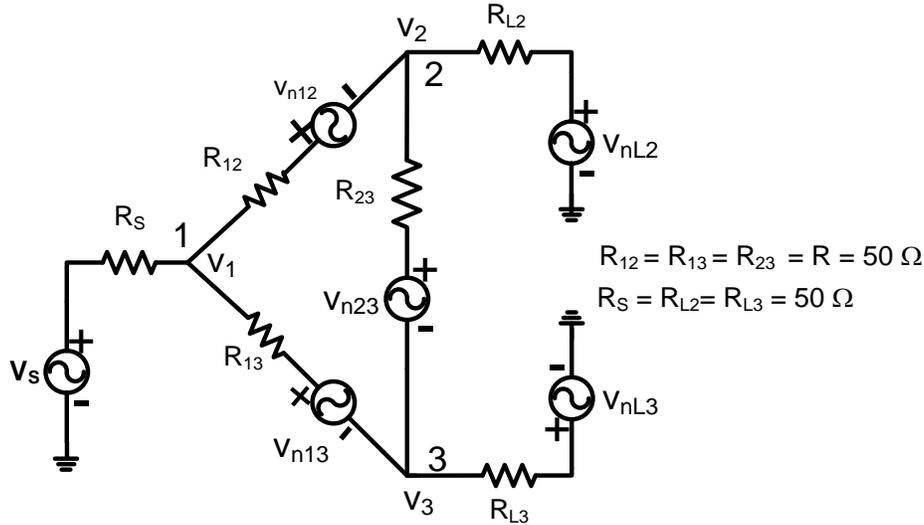

Fig. 6: Equivalent circuit of the delta 3-R power splitter including thermal noise sources for each resistor.

The expectation of the cross-PSD between two output signals $v_2$ and $v_3$ is represented by

$$\langle S_{23}(f) \rangle = \frac{1}{T} \langle V_2(f) V_3^*(f) \rangle$$

$$= \frac{1}{T} \left[ \frac{1}{16} \left( V_S V_S^* - V_{n23} V_{n23}^* + 2V_{nL2} V_{nL2}^* + 2V_{nL3} V_{nL3}^* \right) \right], \quad (6)$$

where, the Fourier transforms of $v_s$, $v_{n12}$, $v_{n13}$, $v_{n23}$, $v_{nL2}$ and $v_{nL3}$ are represented by the corresponding capitalized variables. From (5) we see that the source noise is present equally and in same phase at the output ports 2 and 3 in contrast to the thermal noise of the resistor $R_{23}$ which appears $180^0$ out of phase between outputs 2 and 3. The same effect occurs for the 100 Ω isolation resistor used in the CWPS. The implication of this in the cross-spectrum (6) is that the expected value of thermal noise of the $R_{23}$ is subtracted from the noise in $V_S$. It is also important to note that due to the limited isolation of the resistive splitter, the thermal noise of the load resistors also appear correlated in both output channels. In a perfect 50 Ω system the noise of the source $V_S$ will cancel out with $V_{n23}$, leaving only the thermal noise of the load resistors in the output cross-spectrum.

## B. Simulation

The propagation of thermal noise in different types of power splitters was simulated in the Advanced Design System (ADS) software. In addition to noise from the source and power splitter, the thermal noise contribution of isolators and load resistors was also analyzed. Each thermal noise source was modeled as a unique single sideband from the carrier. In this way each individual noise source could be observed and its contribution to the final cross-spectrum easily determined. A *Circuit Envelope* method of simulation was chosen to be able to include frequency dependent effects of the complex terminations, reactive splitters, and to enable the inclusion of isolators in the analysis. The contribution of the various circuit noise sources to the cross-spectrum between the two outputs was determined. Simulations for various power splitter configurations were tested. The block diagram for the simulation is shown in Fig. 7.

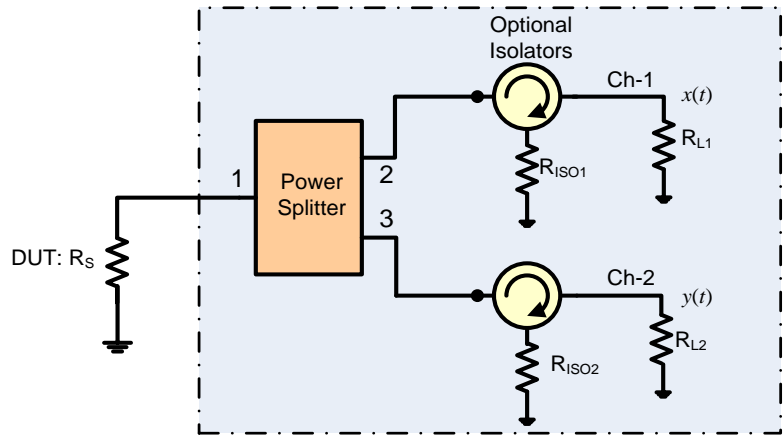

Fig. 7: The main sources of thermal noise used for ADS simulation. $R_s$ represents the thermal noise of the source or the device under test (DUT). Total number of resistors in the power splitter varies from 0 to 3 depending on the configuration. Optional isolators, with thermal resistances are indicated by $R_{ISO1}$ and $R_{ISO2}$. The load resistors $R_{L1}$ and $R_{L2}$ represent the thermal noise of the measurement system.

Table 1 tabulates the results of thermal noise contribution of the individual component to the output cross-spectrum as a fraction of the noise from $R_s$. The simulation is performed for load and source impedances equal to 50 Ω and at 300 K temperature. The values reported in the table are from the expected value of the cross-spectrum. All uncorrelated cross-terms, which reside in the imaginary component of the cross-spectrum, have averaged to zero and the result is an entirely real component.

**Table 1: 2-way Power Splitter (PS): Source impedance ($Z_S$) = Load Impedance ($Z_L$) = 50 Ω, T = 300 K, Isolator: Insertion Loss = 0 dB, Isolation = ∞**

| 1 | 2 | 3 | 4 | | | 5 | 6 | 7 | 8 | 9 | 10 |
|---|---|---|---|---|---|---|---|---|---|---|---|
| Case # | Type of Power splitter (PS) | \multicolumn{5}{c}{Relative cross-spectrum of individual component $S_{ch2-ch1}(f)/S_{R_s}(f)$} | | | | | \multicolumn{2}{c}{Total Noise} |
| | | $R_s$ | \multicolumn{3}{c}{Power Splitter} | ISO#1 | ISO#2 | $R_L$#1 | $R_L$#2 | Without $R_s$ | All Components |
| | | | R1 | R2 | R3 | | | | | | |
| 1 | Wilkinson $R_i$ = 100 Ω | 1 | \multicolumn{3}{c}{-1} | - | - | 0 | 0 | -1 | 0 |
| 2 | Wilkinson $R_i$ = ∞ | 1 | \multicolumn{3}{c}{-} | - | - | -3/2 | -3/2 | -3 | -2 |
| 3 | Wilkinson $R_i$ = ∞, Isolators | 1 | \multicolumn{3}{c}{0} | -0.5 | -0.5 | 0 | 0 | -1 | 0 |
| 4 | 3-R Wye $R_x$ = ~17 Ω | 1 | 1/3 | -2/3 | -2/3 | - | - | 2 | 2 | 3 | 4 |
| 5 | 3-R Wye, $R_x$ = ~17 Ω, Isolators | 1 | 1/3 | -2/3 | -2/3 | 0 | 0 | 0 | 0 | -1 | 0 |
| 6 | 3-R Delta $R_x$ = 50 Ω | 1 | 0 | -1 | 0 | - | - | 2 | 2 | 3 | 4 |
| 7 | 3-R Delta, $R_x$ = 50 Ω Isolators | 1 | 0 | -1 | 0 | 0 | 0 | 0 | 0 | -1 | 0 |
| 8 | 2-R $R_x$ = 50 Ω | 1 | 0 | -3/4 | -3/4 | - | - | 5/4 | 5/4 | 1 | 2 |
| 9 | 2R, $R_x$ = 50 Ω, Isolator | 1 | 0 | -3/4 | -3/4 | 1/4 | 1/4 | 0 | 0 | -1 | 0 |
| 10 | 1-R, R = 25 Ω | 1 | 1/2 | - | - | - | - | 9/5 | 9/5 | 5 | 6 |
| 11 | 1-R, R = 25 Ω, Isolator | 1 | 1/2 | - | - | -3/4 | -3/4 | 0 | 0 | -1 | 0 |

Here, $R_i$ and $R_x$ respectively correspond to the isolation resistor and the resistors for 2-R and 3-R power splitters.

Referring to Table 1, column 1, let us first consider case #1 for the CWPS where the isolating resistor ($R_i$) is 100 Ω. The power splitter noise ($R_i$, column 4) is equal in magnitude to the source noise ($R_s$, column 3) but $180^0$ out of phase. A negative real portion of the cross-spectrum indicates an anti-correlated cross-spectrum. Columns 5 and 6 are blank because no isolators were used for this case. The noise contribution of the load resistors in columns 7 and 8 is zero due to the large isolation between two output ports of the CWPS. The summed noise contribution of all the individual components except $R_s$ is presented in column 9. For case #1, the noise of the isolating resistor ($R_i$) is equal and anti-correlated (indicated by a negative real quantity) to the $R_s$ source resistor noise. In the final column 10 the total noise of all components is shown. The total noise of all components is zero for case #1, a clear indication of a complete cross-spectrum collapse. Note that all the power splitting configurations shown in Table 1 exhibit either a complete cross-spectral collapse or are limited by the noise of the load resistances: none can measure the noise of $R_s$.

Additional simulations, with realistic isolator parameters (isolation of 10 to 30 dB, instead of ∞) produced various intermediate levels of partial correlation collapse. The results from these simulations for various power splitters are shown in Table 2.

**Table 2: 2-way Power Splitter (PS): Source impedance ($Z_S$) = Load Impedance ($Z_L$) = 50 Ω, T = 300 K, Isolator: Insertion Loss = 0.5 dB, Isolation = 15 dB**

| 1 | 2 | 3 | 4 | | | 5 | 6 | 7 | 8 | 9 | 10 |
|---|---|---|---|---|---|---|---|---|---|---|---|
| Case # | Type of Power splitter (PS) | Relative cross-spectrum of individual component $S_{ch2-ch1}(f)/S_{R_s}(f)$ | | | | | | | | Total Noise | |
| | | $R_s$ | Power Splitter | | | ISO#1 | ISO#2 | $R_L$#1 | $R_L$#2 | Without $R_s$ | All Components |
| | | | R1 | R2 | R3 | | | | | | |
| 1 | Wilkinson $R_i = \infty$, Isolators | 1 | - | | | -0.62 | -0.62 | -0.21 | -0.21 | -1.66 | -0.66 |
| 2 | 3-R Wye, $R_x = \sim 17$ Ω, Isolators | 1 | 1/3 | -2/3 | -2/3 | 0.35 | 0.35 | 0.38 | 0.38 | 0.46 | 1.46 |
| 3 | 3-R Delta, $R_x = 50$ Ω Isolators | 1 | 0 | -1 | 0 | 0.35 | 0.35 | 0.38 | 0.38 | 0.46 | 1.46 |
| 4 | 2-R, $R_x = 50$ Ω, Isolator | 1 | 0 | -3/4 | -3/4 | 0.40 | 0.40 | 0.20 | 0.20 | -0.30 | 0.70 |
| 5 | 1-R, R = 25 Ω, Isolator | 1 | 1/2 | - | - | -0.14 | -0.14 | 0.54 | 0.54 | -1.30 | 2.30 |

In addition to the power splitters discussed in Table 1, other devices such as directional couplers, $90^0$ and $180^0$ hybrids, and N-way power splitters were tested. They all introduced phase-inverted thermal noise between two channels.

The conclusions of the simulation are as follows:
1. Resistive power splitters
    a. Resistive power splitters do not have sufficient isolation to allow a cross-spectrum measurement to overcome the loss of signal to noise in each individual channel. They cannot be used to accurately measure a thermally limited source because the dominating noise of the load to the power splitter appears correlated in both channels and cannot be rejected.
    b. 3-R (Delta or Wye) and 2-R splitters produce anti-correlated thermal noise

between the outputs.
2. Reactive splitters (Wilkinson, Hybrid-90, Hybrid-180, couplers)
    a. The isolation resistor produces anti-correlated thermal noise which is equal in magnitude to that of the source resistor.
    b. While removal of the isolation resistor (i.e., MWPS) eliminates the anti-correlated thermal noise, it destroys the isolation and prevents the measurement of a thermally limited source.
3. Isolators, which are circulators with the third port terminated with 50 Ω, present the thermal noise of the isolating resistor at their input. This makes them essentially useless for improving the performance of a splitter with low isolation in terms of thermal noise.
    a. Patching any of the splitters configurations (MWPS, 3-R, 2-R or 1-R) with an ideal isolator produces a complete thermal noise correlation collapse.
    b. Simulation with realistic isolator parameters (isolation of 10 to 30 dB) produces various intermediate levels of partial correlation collapse.

*C. Experimental Results*

For the experimental verification we chose an ultra-low-phase noise oscillator at 100 MHz. The schematic of the oscillator is shown in Fig. 8a, it contains a high-Q clean-up filter and 3 dB attenuator at the output. The output impedance of this oscillator is frequency dependent, it presents 50 Ω at the resonant frequency but a non-50 Ω impedance at Fourier frequencies away from the resonance [22]. Measured s-parameter $|S_{11}|$ and the smith chart for this oscillator are also shown in Fig. 8b and c.

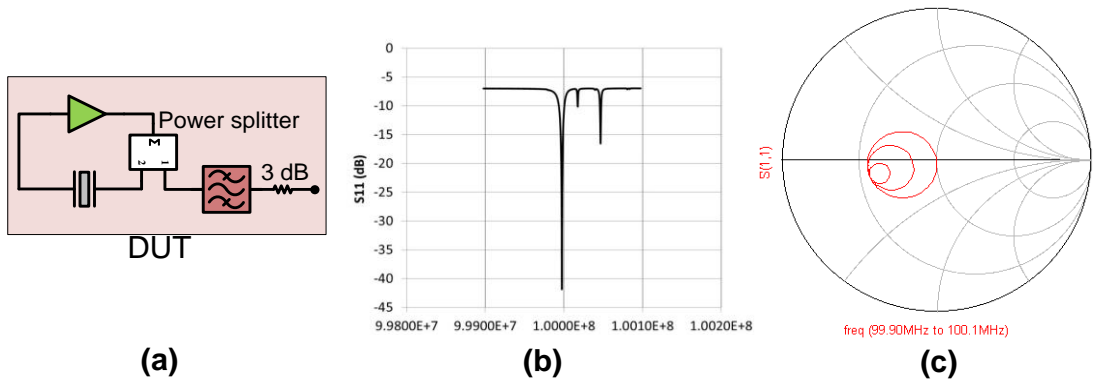

(a)  (b)  (c)

Fig. 8: (a) Schematic of the 100 MHz oscillator with cleanup filter, (b) Measured data of $|S_{11}|$ in dB and (c) Smith chart.

The measurement set-up as shown in Fig. 3 is used for the phase noise measurement of this oscillator. A variable dc offset voltage was added at the input of the PLL integrator to optimize the rejection of the DUT AM noise. With a few exceptions, the AM noise of the DUT was rejected by more than 30 dB to minimize the effect of anti-correlation collapse due to the AM noise leakage. The phase noise of the oscillator was first measured with a CWPS. Assuming a 50 Ω system, and taking into account the DUT power loss in the impedance matching and harmonic filtering (IMHF) circuit in the common path, the theoretical noise should be -189.5 dBc/Hz i.e., -177 - $P_{PS}$. As shown in Fig. 9, a complete collapse (limited by the number of FFT averages) of the noise spectrum was observed due to the anti-correlated thermal noise of the CWPS. Initially, it was thought that the problem of thermal noise of the CWPS could be resolved by removing the 100 Ω isolating resistor and the required isolation restored with the introduction of ferrite isolators at the outputs of the power divider. This modification of the power splitter is represented as MWPS and shown in Fig. 5b. The MWPS does provide good isolation between port 2 and 3, however, the thermal noise of the 50 Ω termination of these isolators appears via port 1 and a ½-wave transmission line to the other channel again causing an anti-correlation collapse. The phase relation of the isolators between two channels can be seen in case #3 of Table 1 and the resulting measured PM noise of the DUT affected by the anti-correlated thermal noise of the MWPS is shown in Fig. 9.

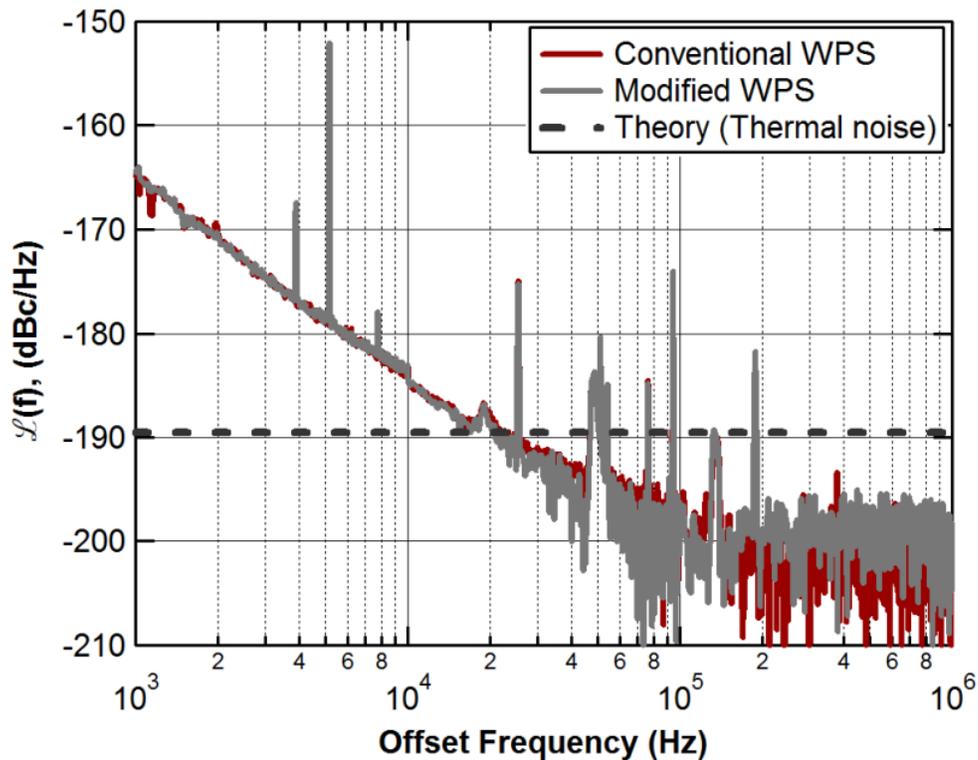

Fig. 9: Phase noise of a 100 MHz oscillator measured with a conventional Wilkinson power splitter (CWPS) and a modified WPS. Theoretical noise of this oscillator referenced to the input power of common-mode power splitter, $P_{PS}$ is -189.5 dBc/Hz calculated from (-177 – $P_{PS}$) assuming a 50 Ω system. The far-from-the-carrier noise in both cases are limited by the maximum FFT number, N = 100,000 available on the analyzer but there is clear indication of a spectrum collapse.

We also measured the phase noise of the same 100 MHz oscillator using resistive 1-R, 2-R and 3-R (Wye configuration) power splitters. We observed large variations in the measured phase noise. For each splitter type, three different measurement configurations were used: (*a*) a direct connection between power splitter and the phase detector, (*b*) a ferrite isolator was introduced between the power splitter and the phase detector in each channel, and (*c*) the isolators were replaced with amplifiers (as shown in the inset of Fig. 10). For configurations (*a*) and (*b*), the measured and the simulated thermal phase noise did not agree because the simulations were performed with an ideal 50 Ω load impedance. However, in actual practice the power splitter is connected to the reactive load of the double balanced mixer used as a phase detector. When the isolator is replaced by an amplifier (*c*) it provides higher isolation and a better impedance match out of the power splitter. This configuration resulted in a closer agreement between the simulation and the experimental results. Fig. 10 shows the experimental result of phase noise measured with amplifiers as well as a strong anti-correlation collapse limited only by the number of FFT averages. It is also observed that the amount of anti-correlation collapse increases with higher isolation between the power splitter and the phase detector.

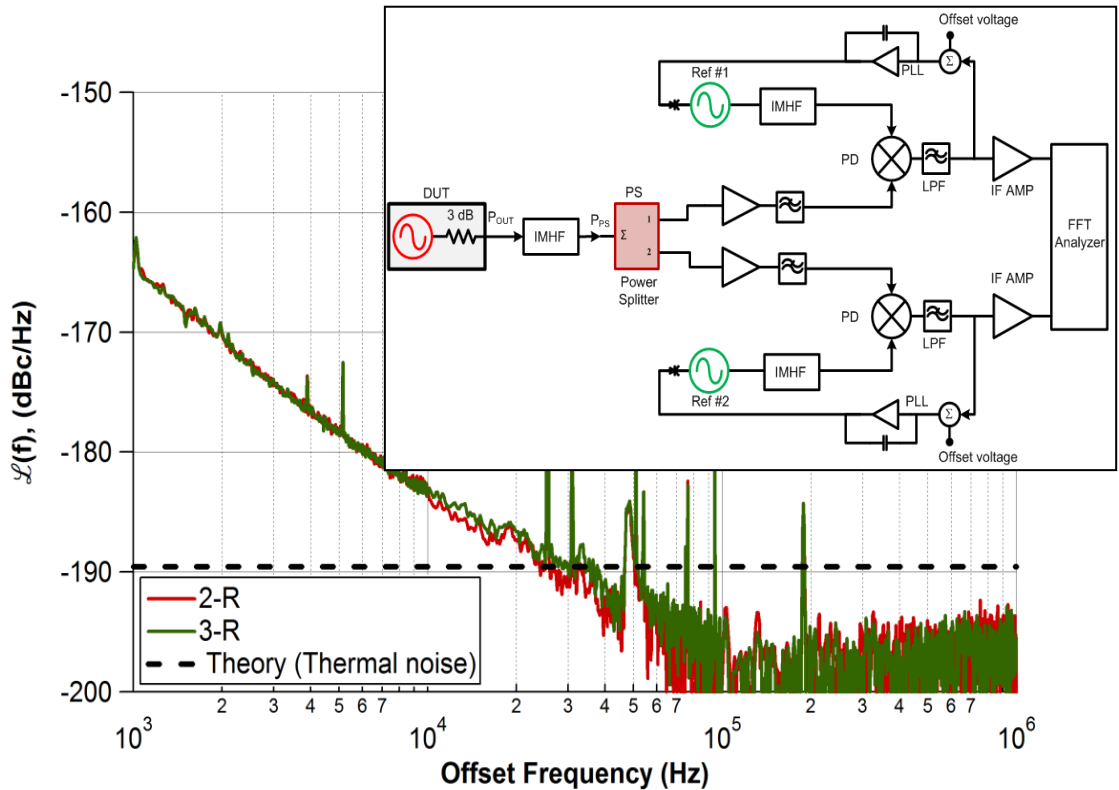

Fig. 10: PM noise of the 100 MHz oscillator measured with resistive 2-R and 3-R power splitters. The measurement configuration is depicted in the inset. The measured thermal phase noise is average limited but again there is an indication of noise spectrum collapse due to the anti-correlation effect. The measured noise is significantly lower than the theoretical thermal noise.

Similar tests were performed for AM measurements of the same oscillator using the configuration shown in Fig. 11. The power splitter is directly connected to an AM detector in each channel whose input impedance is almost a perfect 50 Ω. AM noise was measured with resistive and reactive power splitters and the thermal AM matches the simulation results for each power splitter as shown in Fig. 12. The simulation results correspond to case #10, #8, #4, and #1 in Table 1. For resistive splitters there is a positive-correlation and the noise is higher than the theoretical thermal noise. This is due to the lack of isolation between the AM detectors' input impedance noises. On the other hand, the AM noise measured with the CWPS leads to an anti-correlation collapse as expected from the simulation.

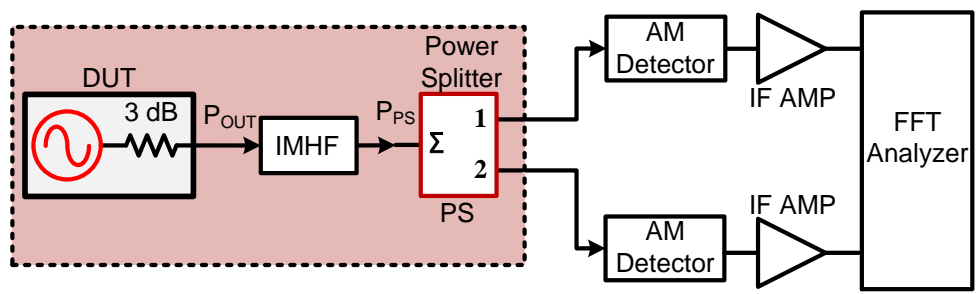

Fig. 11: Block diagram of a dual-channel cross-spectrum system used for measuring AM noise of the DUT. IMHF – Impedance matching and harmonic filtering

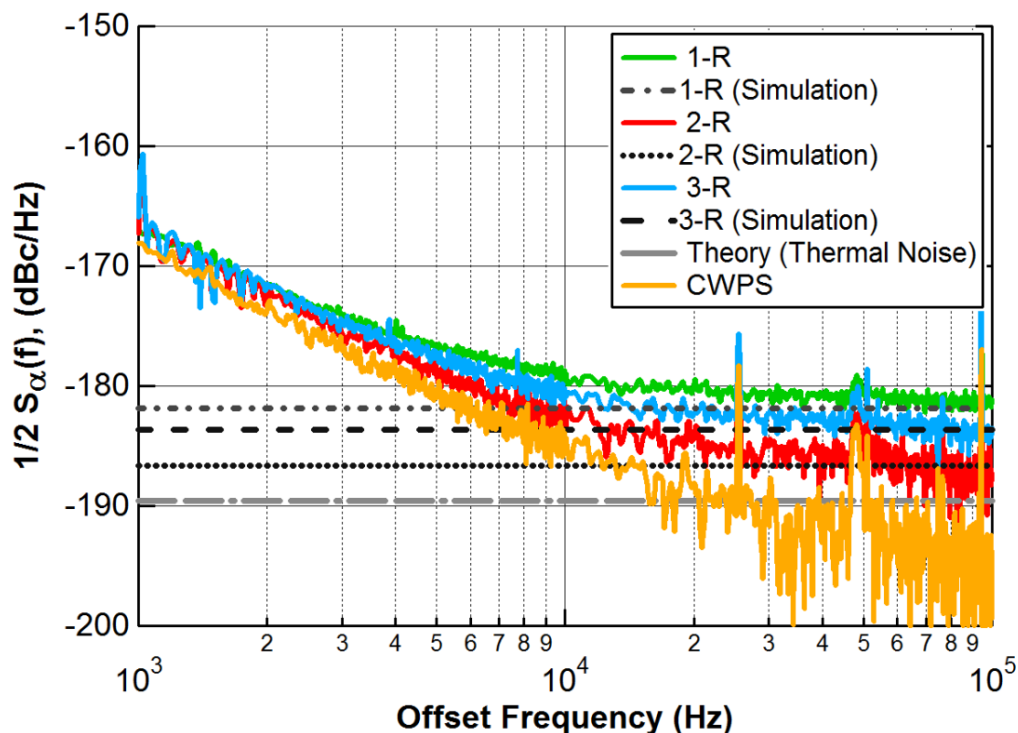

Fig. 12: AM noise of the thermally limited 100 MHz oscillator measured with resistive 1-R, 2-R and 3-R and Wilkinson power splitters. There is a close agreement between experiment and the simulation results. The simulation results correspond to case # 10, #8, #4 and #1 in Table 1.

## V. Summary

We discussed the difficult challenge of PM and AM noise measurement of oscillators at or near the thermal limit of the source impedance and also discussed the limitations of the cross-spectrum system widely used for such measurements. Our conclusions from different simulation and experimental results are as follows:

1. While reactive power splitters such as the Wilkinson have sufficient isolation to measure the thermal noise of source resistance ($R_s$), the thermal noise of the isolation resistor ($R_i$) appears anti-correlated and is subtracted from the $R_s$ noise in the cross-spectrum. This produces a complete collapse in a perfect 50 Ω system.
2. Resistive power splitters do not have sufficient isolation to allow a cross-spectrum measurement to overcome the loss of signal to noise in each individual channel. They cannot be used to accurately measure a thermally-limited source because the dominating noise of the load to the power splitter appears correlated in both channels.
3. Patching any of the low-isolation splitters configurations (MWPS, 3-R, 2-R or 1-R) with an ideal isolator also produces a complete correlation collapse of thermal noise. Simulation with realistic isolator parameters (isolation of 10 to 30 dB) produces various intermediate levels of partial correlation collapse.
4. In practical measurements, the delicate balance between correlated terms and anti-correlated terms that cause these partial or complete collapses are subject to environmental and circuit variations that make the measurement of noise near the thermal limit of $R_s$ extremely difficult to do with any confidence.

In conclusion, all room temperature power splitter configurations we tested, reactive or resistive, introduce either positive or negative correlation biases for heterodyne cross-spectrum measurements near the thermal limit. Any measurement within 10 dB of the thermal limit will have significant bias.

One possible solution to mitigate this problem is to cool the power splitter to cryogenic temperatures. If the power splitter is cooled to a liquid helium temperature (4 K), then its thermal noise will decrease by 19 dB compared to room temperature (300 K). In the near future, we will test a cryogenic Wilkinson splitter to measure the noise of an ultra-low-thermal- noise limited oscillator. However, the non-50 Ω output impedance of such oscillators may cause problems in that they degrade the isolation of the Wilkinson splitter and may cause measurement limitations even when the isolation resistor noise is eliminated. We also will perform a similar analysis on power splitter configurations for residual homodyne methods.

ACKNOWLEDGEMENT

Authors thank Joe Gorin of Keysight Technologies for discovering the important aspect of phase inversion (anti-correlation) in the power splitters that helped to explain the unrealistically low noise level of the thermally limited oscillators.